\documentclass[journal=aamick,manuscript=article]{achemso}
\usepackage[version=3]{mhchem}
\usepackage{amsmath, amssymb, amsthm, bm}

\usepackage{standalone}
\usepackage{lineno}
\usepackage{comment}

\usepackage{xcolor}

\graphicspath{{fig/}}

\author{Haidi Wang}
\author{Wei Lin}

\author{Weiduo Zhu}

\author{Zhao Chen}

\author{Zhongjun Li}
\email{zjli@hfut.edu.cn}

\author{Xiaofeng Liu}
\email{lxf@hfut.edu.cn}

\affiliation{School of Physics, Hefei University of Technology, Hefei, Anhui 230009}

\title{High-throughput calculations of two-dimensional auxetic \ce{M4X8}  with magnetism, electrocatalysis, and alkali metal battery applications}

\begin{document}


\begin{tocentry}
	
\centering
\includegraphics[width=1.0\textwidth]{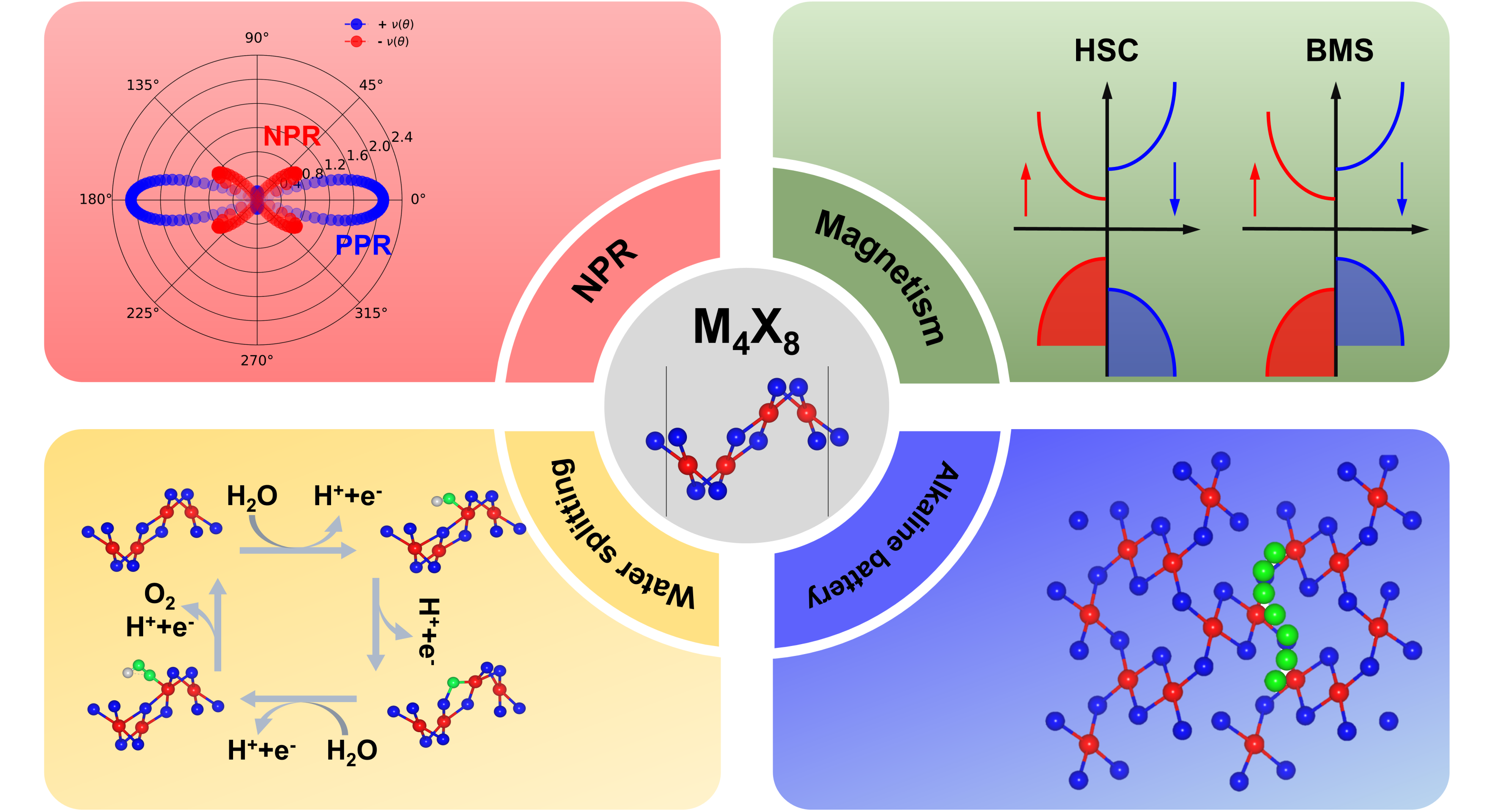} 

\end{tocentry}


\begin{abstract}

Two-dimensional (2D) materials with multifunctional properties, such as negative Poisson's ratio (NPR), magnetism, catalysis, and energy storage capabilities, are of significant interest for advanced applications in flexible electronics, spintronics, catalysis, and lithium-ion batteries. However, the discovery of such materials, particularly in low-dimensional forms, remains a challenge. In this study, we perform high-throughput density-functional theory (DFT) calculations to explore a new class of 2D V-shaped monolayers with remarkable physicochemical properties. Among 18 stable \ce{M4X8} (M = transition metal; X = halogen) compounds, we identify 9 auxetic monolayers, with \ce{Pd4I8} standing out for its exceptionally high NPR of -0.798. Notably, 4 of these materials exhibit half semiconductor properties, while 5 others are bipolar magnetic semiconductors, offering a unique combination of electronic and magnetic behavior. Additionally, these materials demonstrate promising catalytic activity for hydrogen and oxygen evolution reactions (HER/OER) and show potential as anodes for rechargeable metal-ion batteries, particularly in alkali-ion systems. This work not only expands the family of 2D NPR materials but also introduces new candidates with multifunctional capabilities for a wide range of applications in nanoelectronics, catalysis, and energy storage.

\end{abstract}


\section{Introduction}

In recent years, two-dimensional (2D) materials with multi functionalities, such as those exhibiting negative Poisson's ratio (NPR), magnetism, and electrocatalytic properties, have significantly expanded their application prospects in fields like flexible electronics, devices, and energy conversion. These advancements have made them a focal point of research. Among them, 2D NPR materials have gained increasing attention due to their potential applications in stress regulation and energy absorption.\cite{greaves2011poisson} For instance, the inherent auxetic properties of MXene\cite{wu2018highly} materials and $\delta$-phosphorene\cite{wang2017delta} endow them with significant application value in flexible electronics and optical devices. Magnetic 2D materials have also become a research focus due to their potential applications in spintronics. Materials such as \ce{CrI3}\cite{huang2017layer} and \ce{Fe3GeTe2}\cite{deng2018gate} exhibit stable ferromagnetism and hold promise for the development of spin valves and storage devices.\cite{song2019switching} Additionally, 2D materials have achieved remarkable progress in the field of catalysis, leveraging their high specific surface area, excellent conductivity, superior electronic transport properties, and rich surface chemical activity. For example, \ce{MoS2}\cite{chen2023situ} exhibits high conductivity and chemical stability, with its nanosheet structure providing abundant active sites, thereby demonstrating excellent performance in both the hydrogen evolution reaction (HER) and the oxygen evolution reaction (OER).

The discovery and development of new 2D materials, particularly multifunctional 2D materials, remain highly challenging research topics. To address this, various strategies for designing 2D materials have been proposed, including Wyckoff position extension,\cite{evarestov2017use} global optimization,\cite{shi2023accessing, ma2024designing} and element substitution et al.\cite{liu2018brief, huang2022searching} Among these, element substitution within the same group stands out as a simple and effective approach, enabling the rapid design of novel structures with desirable functionalities. For example, researchers have discovered 2D materials with unique properties by employing element substitution and lattice engineering in various material systems, such as pentagonal 2D materials (e.g., penta-C,\cite{zhang2015penta} penta-\ce{SiC2}\cite{liu2016disparate}), black phosphorus-like materials (e.g., black arsenic,\cite{zhang2023intralayer} SnSe\cite{zhang2020unusual}), and layered transition metal dichalcogenides (TMDs) resembling \ce{MoS2} (e.g., \ce{MX2}\cite{yu2017negative}). These strategies enable the optimization of material properties and the identification of novel materials with superior performance, advancing the development of functional 2D materials.

In this study, we focus on a special 2D material, the V-shaped \ce{M4X8} (M = transition metal; X = halogen)  Materials. As a typical example, research on bulk-phase \ce{PtI2} dates back to 1986,\cite{thiele1986platiniodide} and \ce{PtI2} has attracted significant attention owing to its multifunctional properties. Recently, Stoppiello et al. synthesized novel \ce{PtI2} and \ce{PtS2} nanostructures,\cite{stoppiello2017one} further expanding their application potential. Furthermore, Shen et al. discovered that the monolayer \ce{PtI2} exhibits excellent optical and catalytic properties,\cite{shen2019single} and it has a relatively low exfoliation energy, making it easier to isolate and thus more feasible for device applications. Based on these characteristics, we further explored derivatives of the V-shaped 2D material based on \ce{PtI2} through element substitution and high-throughput calculations, with the aim of discovering more functional 2D materials with potential applications. The choice of the V-shaped structure was partly based on chemical intuition, as such structures are believed to exhibit NPR properties, and the channels formed by the V-shaped structure may also facilitate ion migration. Furthermore, particular attention was paid to their magnetic properties when designing these novel materials. Given the critical role of magnetic materials in storage, sensing, and spintronics applications, transition metal elements with potential magnetic characteristics were selected as candidates. For instance, elements such as nickel (Ni), known for their excellent magnetic properties, were included in the study. Given the outstanding catalytic performance of \ce{PtI2} in the OER, as demonstrated in previous studies,\cite{shen2019single} we conducted element substitution with materials possessing excellent catalytic properties, such as cobalt (Co) and palladium (Pd), to further explore the catalytic potential of V-shaped materials.

\section{Computational Methods}
First-principles calculations were performed using the Vienna Ab Initio Simulation Package (VASP)\cite{kresse1993ab, kresse1993abi, kresse1996efficiency} with a plane-wave basis set and the projector augmented-wave (PAW)\cite{kresse1999ultrasoft} method. The exchange-correlation interactions were described using the Perdew-Burke-Ernzerhof (PBE) functional within the generalized gradient approximation (GGA).\cite{perdew1996generalized} A 500 eV energy cutoff was used, and a vacuum slab of over 20 $\text{\AA}$ was included to avoid spurious interactions between periodic images. The atomic positions and lattice constants were relaxed using the conjugate gradient method. The convergence criterion for total energy was set to $10^{-5}$ $\text{eV}$, and the residual forces on each atom were converged to within 0.01 $\text{eV/\AA}$.

We employed a 9$\times$7$\times$1 k-point mesh using the Monkhorst–Pack (MP)\cite{monkhorst1976special} scheme to sample the Brillouin zone of the unit cell. Since the PBE functional underestimates the band gap of semiconductors, the Heyd–Scuseria–Ernzerhof (HSE06)\cite{heyd2003hybrid} hybrid functional was used for band gap correction. To qualitatively assess the structural stability, we examined the thermodynamic stability through cohesive energy calculations. The dynamical stability was evaluated using phonon dispersion calculations via the Phonopy code,\cite{togo2015first} based on the finite difference method implemented in VASP. Additionally, ab initio molecular dynamics (AIMD) simulations were performed using a 3$\times$3$\times$1 monolayer supercell to assess thermal stability at 300 K for a simulation time of 5 ps with a time step of 1 fs. Finally, the elastic constants of the monolayer \ce{M4X8} were calculated using the mech2d\cite{wang2023mech2d} software package to verify the mechanical stability and to analyze the Young's modulus and Poisson's ratio. Finally, the alkali ion diffusion barriers in monolayer \ce{M4X8} were estimated by using the climbing image nudged elastic band (CI-NEB) method.\cite{hen2000climb}
 
\section{Results and Discussion}

\subsection{Structure and Stability }

\begin{figure}[H]
\centering
\includegraphics[width=1.0\textwidth]{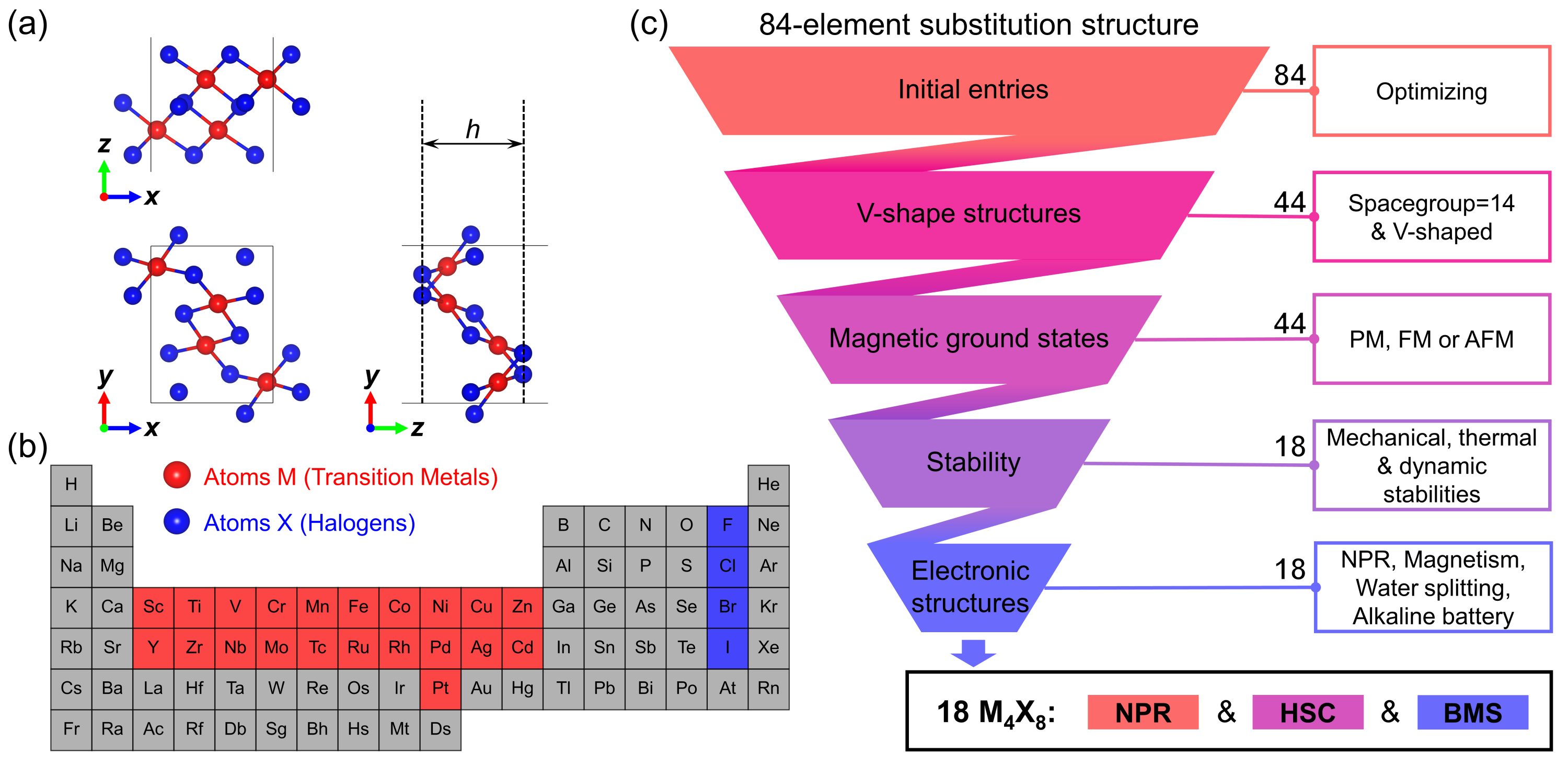}
\caption{(a) Three-view of \ce{M4X8}, with red spheres representing metallic element M and blue spheres representing element X. M and X represent transition metal atoms and halogen atoms. (b) Elements used for substitution. (c) Screening workflow for \ce{M4X8} and the number of remaining structures at each step.}
\label{fig:structure}
\end{figure}

As illustrated in Figure \ref{fig:structure}(a), the primitive cell of the \ce{M4X8} monolayer includes four M atoms (depicted as red spheres) and eight X atoms (depicted as blue spheres), forming a rectangular lattice. Each M atom is bonded to four X atoms, while each X atom is coordinated with two M atoms. From a side view, the \ce{M4X8} monolayer exhibits a zigzag-shaped, V-like corrugated structure with a buckling height of $h$. This arrangement lacks mirror symmetry but maintains centro-symmetry. The \ce{M4X8} monolayer inherits the non-metallic shielding characteristics of traditional transition metal dichalcogenides (TMCs), such as the 1T and 2H phases, with metal M atoms situated in the central layer and non-metallic X atoms positioned in the outer layers.

Based on the \ce{PtI2} monolayer with $P2_1/c$ (No.14) symmetry, we constructed 84 V-shaped corrugated monolayers by substituting Pt and I atoms with other elements of the periodic table. Figure \ref{fig:structure}(b) highlights the selected elements: M includes 21 transition metals (Sc-Zn, Y-Cd, and Pt), and X includes 4 halogens (F, Cl, Br, and I). Figure \ref{fig:structure}(c) illustrates the workflow of our high-throughput calculations. Among the 84 optimized \ce{M4X8} monolayers, 40 structures either lost their corrugated configuration or failed to maintain $P2_1/c$ symmetry, indicating that these structures are unsuitable for the V-shaped corrugated configuration. We focus on the remaining 44 monolayers, whose optimized structures retain the initial symmetry and potentially exhibit NPR characteristics. To further determine the magnetic ground states of these 44 structures, we considered not only paramagnetic and ferromagnetic states but also three types of antiferromagnetic (AFM) states, with specific spin arrangements detailed in Supplementary Figure S1. The calculations reveal that even combinations of the same metal element with different halogens can exhibit distinct magnetic ground states. For instance, \ce{Cr4Cl8} adopts an AFM state, while \ce{Cr4Br8} and \ce{Cr4I8} exhibit ferromagnetic states.

To evaluate the stability of the proposed materials, we firstly calculated the cohesive energy ($E_{\mathrm{coh}}$) of the \ce{M4X8} structures to assess their thermal stability. The cohesive energy is defined as:
\begin{equation}
E_{\mathrm{coh}} = \frac{x E_{\mathrm{M}} + y E_{\mathrm{X}} - E_{\mathrm{total}}}{x + y}
\end{equation}
where $E_{\mathrm{M}}$, $E_{\mathrm{X}}$, and $E_{\mathrm{total}}$ represent the energy of a single M atom, a single X atom, and the total energy of the \ce{M4X8} unit cell, respectively. The quantities $x$ and $y$ denote the number of M and X atoms in the unit cell. The results show that the cohesive energies of all 44 structures are positive, indicating that these V-shaped materials exhibit high thermodynamic stability and are likely to be synthesizable. We found that the cohesive energy of \ce{M4X8}, formed by the same metal element and different halogens, decreases as the atomic number of the halogen increases. This may be attributed to the decrease in electro-negativity of the X atom, which enhances the ionic character of the metal-halogen bond and weakens its covalent character, leading to a more loosely packed structure of the compound and consequently a reduction in cohesive energy. The basic geometric parameters of these 44 structures, including lattice constants $a$ and $b$, magnetic ground states, and cohesive energies, are summarized in Table S1 of the supplementary material.

To further verify the stability of the \ce{M4X8} monolayers, we investigated their dynamic stability by calculating the phonon dispersion. Among the 44 structures, 22 were found to be dynamically stable (detailed results are provided in Figures S2 and S4). Notably, \ce{Ni4F8} and \ce{Ag4F8} exhibited minor negative frequencies near the $\Gamma$ point (-0.095 and -0.081 THz, respectively), a phenomenon commonly observed in various two-dimensional materials, which does not compromise their dynamic stability. In contrast, the remaining 22 structures with significant negative frequencies were considered dynamically unstable. Subsequently, the thermal stability of the remaining 22 \ce{M4X8} monolayers was assessed via AIMD simulations. As shown in Figure S4, \ce{Ru4Br8}, \ce{Rh4Cl8}, \ce{Tc4Cl8}, and \ce{Tc4Br8} underwent significant structural distortions at 300 K, indicating a lack of thermal stability. In contrast, the total energy fluctuations of the other 18 \ce{M4X8} monolayers remained below 3 eV (Figure S3), and their structures remained stable during the simulations, suggesting good thermal stability. Finally, the mechanical stability of these 18 thermally stable structures was analyzed using the Born criteria ($C_{11}C_{22} - C_{12}{ }^2 > 0$ and $C_{66} > 0$),\cite{ding2013density} where $C_\text{ij}$ denotes the elastic constants. Excitingly, the stability assessments reveal that 18 V-shaped \ce{M4X8} monolayers exhibit excellent thermodynamic, dynamic, thermal, and mechanical stabilities, indicating significant potential for experimental synthesis.

\subsection{Mechanical Properties}

\begin{figure}[H]
\centering
\includegraphics[width=0.9\textwidth]{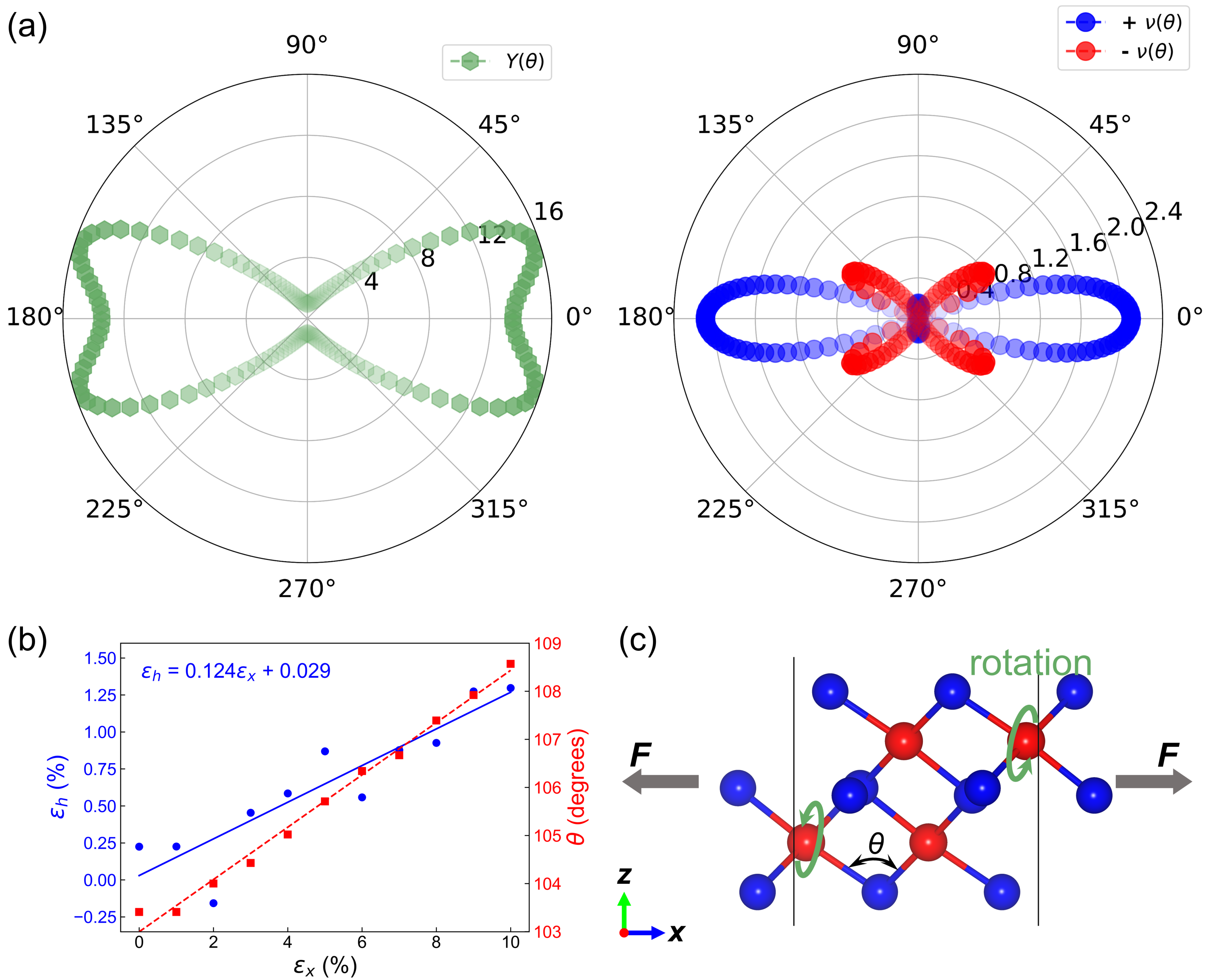}
\caption{(a) The variation of Young's modulus and Poisson's ratio of \ce{Pd4I8} with respect to angle. (b) The strain of the buckling height ($\varepsilon_\text{h}$, \%) and the variation trend of the bond angle ($\theta$, as shown in Figure (c)) of the \ce{Pd4I8} structure under tensile strain along the $x$-axis ($\varepsilon_\text{x}$, \%). Where $\varepsilon_\text{h}$ = 100 $\times$ ($h$ - $h_\text{0}$)/$h_\text{0}$ and $h_\text{0}$ is the buckling height at $\varepsilon_\text{x}$ = 0\%. (c) Schematic illustration of the out-of-plane NPR mechanism.}
\label{fig:npr}
\end{figure}

In the previous section, we examined the basic mechanical properties of 18 stable \ce{M4X8} monolayers. Here, we further discuss their elastic mechanical properties related to elastic constants ($C_{11}$, $C_{12}$, $C_{22}$, and $C_{66}$), such as Young's modulus and Poisson's ratio. Specifically, the angle-dependent Young's modulus $Y(\theta)$ and Poisson's ratio $\nu(\theta)$ can be expressed as follows:\cite{cadelano2010elastic}

\begin{equation}
Y(\theta) = \frac{C_{11} C_{22} - C_{12}{ }^{2}}{C_{11} \sin \theta{ }^{4} + A \sin \theta{ }^{2} \cos \theta{ }^{2} + C_{22} \cos \theta{ }^{4}}
\end{equation}
\begin{equation}
\nu(\theta) = \frac{C_{12} \sin \theta{ }^{4} - B \sin \theta{ }^{2} \cos \theta{ }^{2} + C_{12} \cos \theta{ }^{4}}{C_{11} \sin \theta{ }^{4} + A \sin \theta{ }^{2} \cos \theta{ }^{2} + C_{22} \cos \theta{ }^{4}}
\end{equation}
where $A = \left(C_{11} C_{22} - C_{12}{ }^{2}\right) / C_{66} - 2 C_{12}$, $B = C_{11}+C_{12} - \left(C_{11} C_{22}\right. \left. - C_{12}{ }^{2}\right) / C_{66}$.

Based on the above equations, we present the variation of Young's modulus ($Y$) and Poisson's ratio ($\nu$) with respect to angle ($\theta$) in Figure \ref{fig:npr}(a) (taking \ce{Pd4I8} as an example). The results show that the monolayer \ce{Pd4I8} exhibits NPR characteristics, with a maximum Young's modulus of 15 N/m and a minimum negative Poisson's ratio of -0.798, occurring at $\theta$ = 42°. Table \ref{table:Mechanical} lists the relevant parameters of the mechanical properties for 18 \ce{M4X8} monolayers, all of which exhibit relatively low in-plane Young's modulus (less than 31 N/m) and significant anisotropy (see Figures S5 and S6). The highest in-plane Young's modulus typically occurs along the $x$-axis, which is also the direction with the maximum positive Poisson's ratio. The Young's modulus of these materials is significantly lower than that of graphene (340 N/m)\cite{lee2008measurement} and \ce{MoS2} (330 N/m),\cite{castellanos2012elastic} indicating their promising potential for applications in flexible devices.

Notably, 8 \ce{M4X8} monolayers exhibit NPR, with the lowest NPR values occurring in the direction between $34^\circ$ and $43^\circ$, indicating their excellent mechanical properties such as dent resistance, high fracture toughness, and effective vibration or sound absorption capabilities. Specifically, the NPR values of \ce{Cu4Br8}, \ce{Mo4Br8}, and \ce{Ag4Br8} (-0.257, -0.068, and -0.151, respectively, as shown in Figure S6) are more negative than those of black
phosphorus (-0.027)\cite{jiang2014negative} and $S$-\ce{SiS2} (-0.054)\cite{liu2023computational}, suggesting that these V-shaped monolayers demonstrate more pronounced responses as auxetic materials. Additionally, the V-shaped buckling structures also exhibit out-of-plane NPR characteristics, including \ce{Ag4Br8}, \ce{Cu4Cl8}, \ce{Cu4Br8}, \ce{Mo4Br8}, \ce{Ni4Cl8}, \ce{Pd4Br8}, \ce{Pd4I8}, \ce{Pt4Br8} and \ce{Pt4I8}. As illustrated in Figure \ref{fig:npr}(b), the strain degree of the buckling height ($\varepsilon_\text{h}$) of the \ce{Pd4I8} structure increases with tensile strain applied along the $x$-direction ($\varepsilon_\text{x}$). Based on the linear fitting, the out-of-plane NPR of the \ce{Pd4I8} structure is predicted to be -0.124. To explain the mechanism of out-of-plane NPR, Figure \ref{fig:npr}(c) illustrates the local structure. As can be seen, applying tensile strain along the $x$-axis causes the lattice constant $a$ to increase, $b$ to decrease, and the bond angle $\theta$ to expand. This transformation induces a rotation of the \ce{MX4} units within the unit cell, causing the two central \ce{MX4} units to adopt a more vertical alignment. Consequently, the structural buckling height increases, leading to an out-of-plane negative expansion effect.

\begin{table}[hb]
\begin{center}
\caption{Calculated values for the \ce{M4X8} monolayer in various structures: elastic constants ($C_\text{ij}$, N/m), maximum value of Young's modulus ($Y_\text{max}$, N/m), minimum value of the Poisson's ratio ($\nu_\text{min}$), and the angle corresponding to the minimum Poisson's ratio ($\theta_\text{min}$).}
\label{table:Mechanical}
\begin{tabular}{lccccccc}
\hline
System      &$C_{11}$&$C_{12}$&$C_{22}$&$C_{66}$&$Y_\text{max}$& $\nu_\text{min}$ & $\theta_\text{min}$ \\
\hline
\ce{Cr4Cl8} & 20.70 & 2.66 &  9.25 & 6.32 & 19.94 &  0.082  & $49.13^\circ$ \\
\ce{Cr4Br8} & 20.26 & 3.61 & 10.86 & 6.24 & 19.06 &  0.154  & $54.15^\circ$ \\
\ce{Cr4I8}  & 18.38 & 3.04 &  6.02 & 5.49 & 16.85 &  0.004  & $45.13^\circ$ \\
			
\ce{Ni4F8}  & 40.02 & 7.66 & 40.02 & 8.22 & 38.55 &  0.191  & $0.00^\circ$     \\  
\ce{Ni4Cl8} & 31.68 & 2.69 &  9.52 & 6.16 & 30.91 &  0.085  & $89.25^\circ$ \\
\ce{Ni4Br8} & 26.19 & 3.14 &  5.62 & 5.74 & 24.43 & -0.033  & $43.66^\circ$ \\
\ce{Ni4I8}  & 25.90 & 2.99 &  7.76 & 6.82 & 24.75 &  0.008  & $45.13^\circ$ \\
			
\ce{Cu4F8}  & 31.28 & 5.82 &  7.43 & 6.30 & 26.72 &  0.073  & $48.13^\circ$ \\
\ce{Cu4Cl8} & 23.38 & 2.20 &  6.26 & 5.52 & 22.61 &  0.007  & $45.13^\circ$ \\
\ce{Cu4Br8} & 18.93 & 1.84 &  2.41 & 3.76 & 17.51 & -0.257  & $38.13^\circ$ \\
			
\ce{Mo4Br8} & 15.78 & 2.51 &  2.57 & 2.93 & 13.32 & -0.068  & $43.14^\circ$ \\
\ce{Mo4I8}  & 16.11 & 2.46 &  5.14 & 4.74 & 14.93 & -0.004  & $44.73^\circ$ \\
			
\ce{Pd4Br8} & 21.97 & 1.84 &  3.91 & 4.15 & 21.10 & -0.057  & $42.15^\circ$ \\
\ce{Pd4I8}  & 19.74 & 3.00 &  1.44 & 4.17 & 15.63 & -0.798  & $34.64^\circ$ \\
			
\ce{Ag4F8}  & 18.97 & 2.72 &  3.63 & 2.55 & 16.93 &  0.137  & $64.18^\circ$ \\   
\ce{Ag4Br8} & 12.22 & 3.19 &  4.38 & 2.83 &  9.90 &  0.187  & $52.14^\circ$ \\
			
\ce{Pt4Br8} & 24.07 & 2.72 &  2.22 & 4.78 & 20.74 & -0.478  & $35.61^\circ$\\
\ce{Pt4I8}  & 23.23 & 1.83 &  4.86 & 5.21 & 22.55 & -0.079  & $41.51^\circ$\\
			
\hline
\end{tabular}
\end{center}
\end{table}

\subsection{Electronic Properties}

\begin{figure}[H]
\centering
\includegraphics[width=1.0\textwidth]{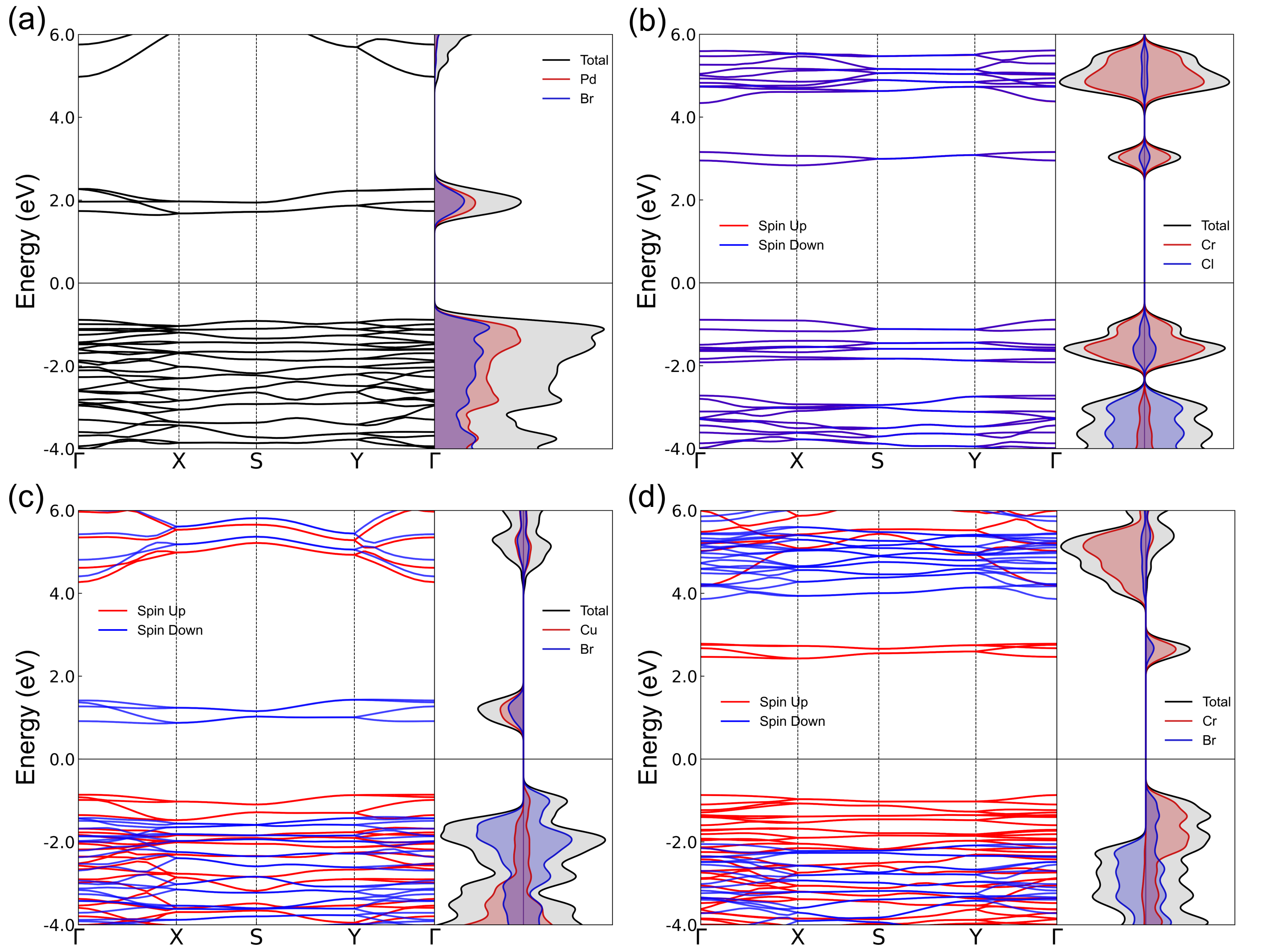}
\caption{Band structure and PDOS of the monolayers: (a) paramagnetic semiconductor \ce{Pd4Br8}; (b) antiferromagnetic semiconductor \ce{Cr4Cl8}; (c) half-semiconductor \ce{Cu4Br8}, and (d) bipolar magnetic semiconductor \ce{Cr4Br8}.}
\label{fig:band}
\end{figure}

Considering the rich diversity of magnetic ground states exhibited by the V-shaped \ce{M4X8} monolayers, we further investigated their electronic properties to expand their potential applications. The electronic band structures (including band gaps and types) and atomic projected density of states (PDOS) for 18 \ce{M4X8} monolayers were calculated (see Table S2). Among these materials, \ce{Ag4Br8} exhibits metallic behavior, while the other 17 structures are semiconductors, all with indirect band gaps. Given that the PBE functional tends to underestimate band gaps, the HSE06 method was employed to obtain more reliable values. The corrected band types remained consistent with the PBE results, though the band gaps of the 17 semiconductors increased to varying extents, while \ce{Ag4Br8} retained its metallicity. We observed that for \ce{M4X8} monolayers composed of the same metal atom but different halogen atoms, the band gap decreases as the atomic number of the halogen increases. The primary reason is probably differences in the atomic radii and electro-negativity of the halogen atoms, which result in variations in the bond lengths and charge distribution between the M and X atoms. Halogens with lower electronegativity (e.g., Iodine) exhibit weaker electron-attracting abilities, raising the energy of the valence band and consequently narrowing the band gap. Moreover, as shown in Table S1, for example, in the case of \ce{Ni4B8}, the lattice constants $a$ and $b$ increase with the atomic number of the halogen. Larger lattice constants generally reduce the ionic character and bond strength of the metal-halogen bonds, resulting in lower transition energies for electrons between the valence and conduction bands, thereby decreasing the band gap.\cite{mosconi2016electronic, chen2021impact}

Through the analysis of band structures and PDOS calculated using the HSE06 functional (as shown in Figure \ref{fig:band} and Table S2), we identified four typical semiconductor types: paramagnetic semiconductors, antiferromagnetic semiconductors, half-semiconductors (HSC),\cite{Li2016NSR, Kimura2003Magnet} and bipolar magnetic semiconductors (BMS),\cite{Li2012Bipolar, Li2013Bipolarmag, Yuan2013Hydrogen, Chen2023Recent, Li2022Areview} For all four categories, the valence band maximum (VBM) and conduction band minimum (CBM) are composed of contributions from the M and X atoms. Excitingly, among the ferromagnetic semiconductors, we identified 2 HSCs (\ce{Cr4Br8} and \ce{Cr4I8}) and 5 BMSs (\ce{Cu4F8}, \ce{Cu4Cl8}, \ce{Cu4Br8}, \ce{Mo4Br8}, and \ce{Mo4I8}). HSCs exhibit semiconducting behavior in one spin channel and semiconducting or insulating behavior in the other. The valence band and conduction band of HSCs are spin-split, with the VBM and CBM having the same spin direction. Due to the complete spin polarization of the VBM and CBM, HSCs can generate 100\% spin-polarized electrons and holes under thermal or optical excitation, or simply through electric gating. Similarly to HSCs, BMSs also allow for electrical control of the spin orientation of charge carriers, a crucial feature for developing high-performance spintronic devices. However, unlike HSCs, the VBM and CBM in BMSs are fully spin-polarized in opposite spin directions. This unique property enables BMSs to control the spin direction of charge carriers by applying positive and negative gate voltages, while also adjusting their conductivity. Based on BMSs, various electrically controlled spintronic devices have been developed, such as bipolar field effect spin filters and field effect spin valves.\cite{li2023nonvolatile} These properties position \ce{Cr4Br8}, \ce{Cr4I8}, \ce{Cu4Br8}, \ce{Cu4Cl8}, \ce{Cu4F8}, \ce{Mo4Br8}, and \ce{Mo4I8} as promising candidates for advancing spintronic technology.

\subsection{Water Splitting}

\begin{figure}[H]
\centering
\includegraphics[width=1.0\textwidth]{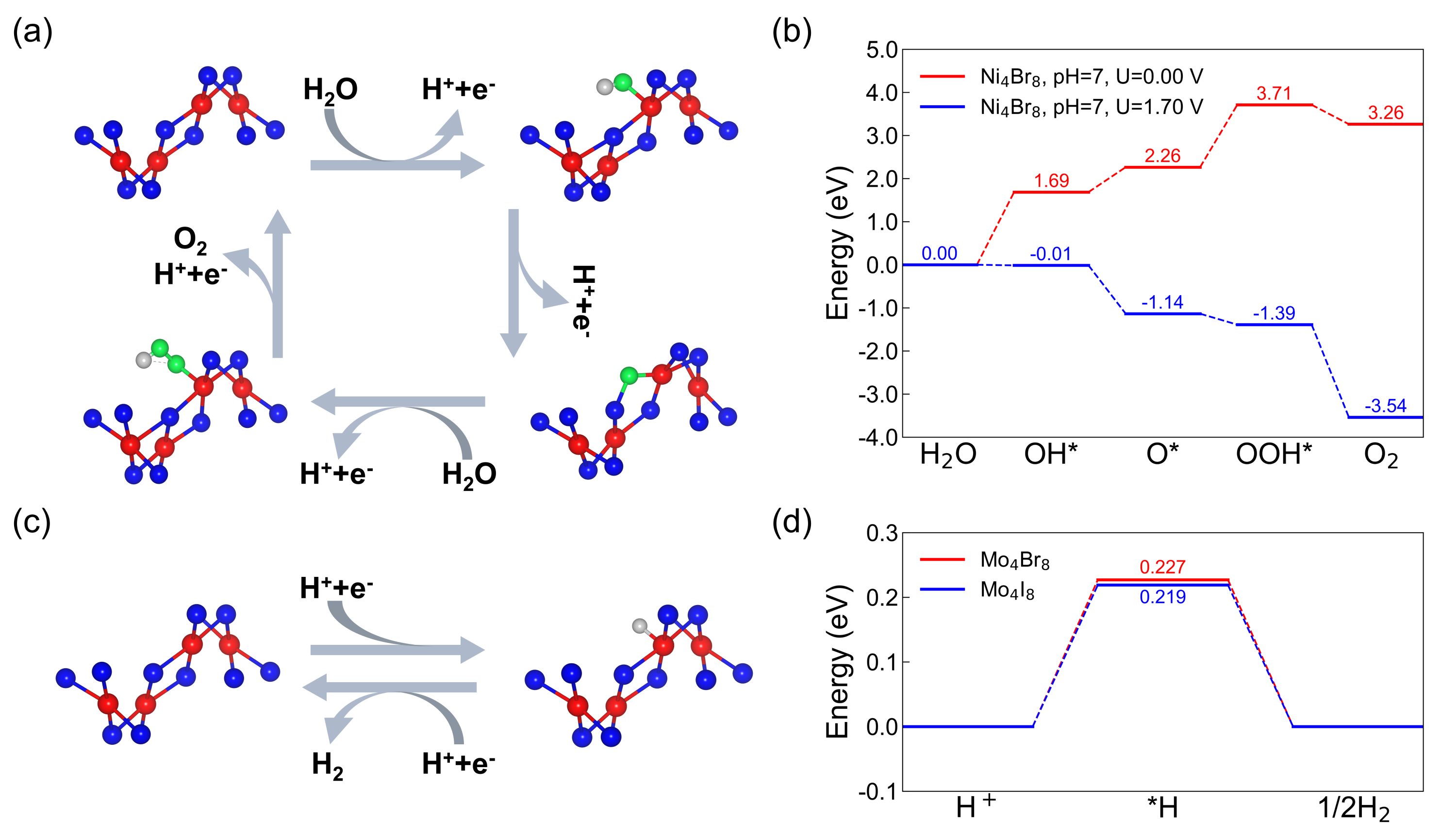}
\caption{(a) Steps of the OER reaction and intermediate reaction structures. (b) $\Delta G$ values for the four steps of the OER reaction on \ce{Ni4Br8}. (c) H$^+$ ion adsorption sites for the HER reaction. (d) $\Delta G$ values for the HER reactions of \ce{Mo4Br8} and \ce{Mo4I8}. In (a) and (c), the green spheres represent O atoms, and the gray spheres represent H atoms.}
\label{fig:heroer}
\end{figure}

As shown in Figures \ref{fig:heroer}(b) and (d), we calculated the OER performance of monolayer \ce{Ni4Br8} and the HER performance of \ce{Mo4Br8} and \ce{Mo4I8}. Figures \ref{fig:heroer}(a) and (c) illustrate the reaction steps and the lowest-energy intermediate structures (OH*, O*, OOH*, H*). For both OER and HER, the adsorption sites of intermediates are preferentially locate on metal atoms. For the OER, which follows a 4e$^-$ reaction pathway, the corresponding free energy (G) profiles are summarized in Figure \ref{fig:heroer}(c). Initially, water adsorbs and transforms into OH*, with $\Delta G$ reaching a maximum of 1.68 eV. Subsequently, OH* is oxidized to O* by releasing H$^+$ and an electron, with $\Delta G$ = 0.578 eV. The O* intermediate is then oxidized to OOH*, with $\Delta G$ = 1.447 eV. Finally, OOH* decomposes into oxygen, protons, and electrons, releasing 0.448 eV of energy. When an external potential (U = 1.70 V) is applied, all OER steps go downhill, indicating that water splitting on \ce{Ni4Br8} under these conditions can occur spontaneously. According to the formula provided in the supplementary information, the over-potential for \ce{Ni4Br8} under pH = 7 conditions is calculated to be 0.870 V. For the HER (Figure \ref{fig:heroer}(d)), \ce{Mo4Br8} and \ce{Mo4I8} also demonstrate excellent performance. Under pH = 0 conditions, their $\Delta G$ values are 0.227 eV and 0.219 eV, respectively. These materials exhibit outstanding catalytic properties, suggesting that further optimization and exploration of their potential in practical applications could be highly promising.

\subsection{Alkali Ions Migration }

\begin{figure}[H]
\centering
\includegraphics[width=0.65\textwidth]{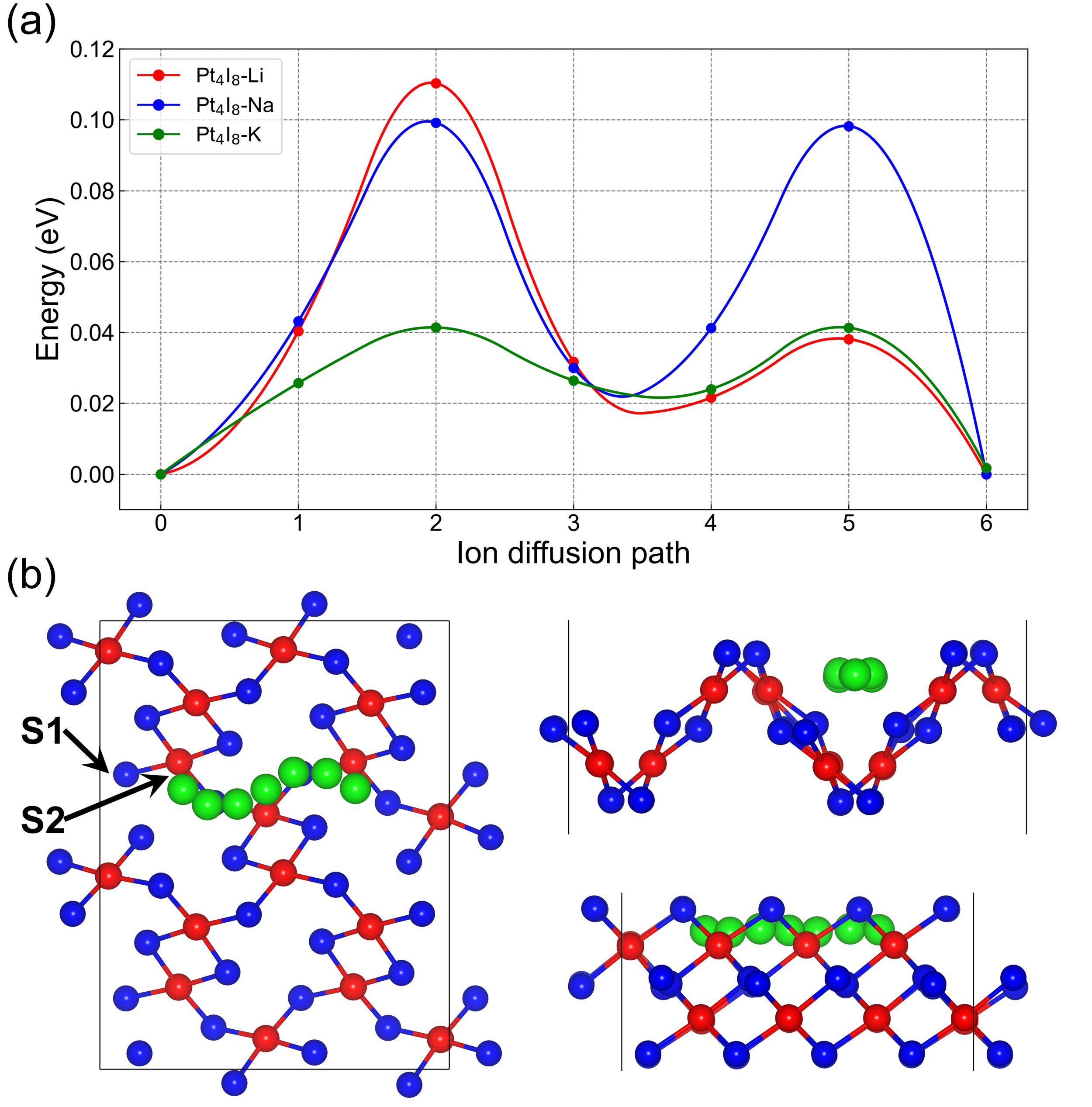}
\caption{(a) The diffusion barrier of Li/Na/K on the \ce{Pt4I8}. (b) The corresponding diffusion paths (taking K ion as an example) on the \ce{Pt4I8}, viewed from three directions.}
\label{fig:neb}
\end{figure}

Building on structural analysis, we hypothesize that the grooves in the V-shaped structure facilitate alkali ion diffusion. Further studies revealed that the \ce{M4X8} supercell forms linear channels along the $x$-direction, which may promote the directional migration of alkali ions. To this end, we calculated the migration properties of a single Li, Na, and K ion in a 2$\times$2$\times$1 monolayer supercell. First, we investigated the adsorption behavior of Li, Na, and K atoms on \ce{M4X8}. Due to the spatial symmetry, only two possible adsorption sites, S1 and S2, were considered, as shown in Figure \ref{fig:neb}(b). The adsorption energy ($E_{\mathrm{ad}}$) was used to determine the most favorable adsorption site. The adsorption energy is calculated using the following formula:
\begin{equation}
	E_{\mathrm{ad}} = E_{\mathrm{\ce{M4X8}-S}} - E_{\mathrm{\ce{M4X8}}} - E_{\mathrm{S}}
\end{equation}
where $E_{\mathrm{\ce{M4X8}-S}}$ is the total energy of the \ce{M4X8} monolayer with the Li/Na/K atom inserted, $E_{\mathrm{\ce{M4X8}}}$ is the total energy of the pristine \ce{M4X8} monolayer, $E_{\mathrm{S}}$ and is the average energy of the Li/Na/K atom in its body-centered cubic (bcc) metallic phase. For the adsorption of Li/Na/K atoms on the \ce{M4X8} monolayer, the $E_{\mathrm{ad}}$ at the S2 site is the lowest, with values of -0.834 eV/atom for Li, -0.890 eV/atom for Na, and -1.440 eV/atom for K. Then the CI-NEB calculations are then used to estimate the diffusion barriers. As expected ( see  in Figure \ref{fig:neb}(a)), when ions migrate along the grooves, the diffusion energy barriers are very small. For Li ions, the diffusion energy barrier is 0.11 eV, which is smaller than that value of green phosphorus (0.14 eV)\cite{wang2022two} and \ce{MoS2} (0.21 eV).\cite{li2012enhanced} For Na ions, the diffusion energy barrier is slightly lower than that of Li, decreasing to less than 0.10 eV, which is still lower than that of \ce{MoS2} (0.28 eV)\cite{mortazavi2014ab} and penta-graphene (0.28 eV).\cite{xiao2016penta} Additionally, the diffusion energy barrier for K ions is only 0.04 eV, which is much lower than that value of \ce{Si3C} (0.18 eV),\cite{wang2020ab} graphite (0.27 eV),\cite{xu2016dispersion} and \ce{TiC3} (0.19 eV),\cite{fatima2024two} and significantly lower than that of Li and Na ions. Moreover, as seen from the corresponding ion diffusion paths in Figure \ref{fig:neb}(b), the migration of the K ion on \ce{M4X8} has minimal impact on the material's structure, demonstrating good structural stability, which is also true for Li and Na. In general, the low diffusion energy barriers indicate that \ce{M4X8} is an excellent platform for ion migration and has potential for use in ion batteries.

\section{Conclusions}

Using DFT calculations, we performed a high-throughput analysis of 84 V-shaped corrugated \ce{M4X8} monolayers. Among these, we identified 18 structures with thermodynamic, dynamic, mechanical, and thermal stability. Furthermore, we explored their potential applications in mechanical systems, spintronics, electrochemical catalysis, and alkali ion battery application. Our study indicates that 9 of the \ce{M4X8} monolayers exhibit NPR characteristics. Notably, \ce{Pd4I8} and \ce{Pt4Br8} demonstrate outstanding NPR values, and their low Young’s modulus makes them promising candidates for flexible devices. The HSC properties of \ce{Cr4Br8} and \ce{Cr4I8}, as well as the BMS characteristics of \ce{Cu4Br8}, \ce{Cu4Cl8}, \ce{Cu4F8}, \ce{Mo4Br8}, and \ce{Mo4I8}, position these materials as promising candidates for spintronic devices. Additionally, \ce{Ni4Br8}, \ce{Ni4Cl8}, \ce{Mo4Br8}, and \ce{Mo4I8} exhibit suitable band gaps and low over-potentials, indicating their potential for photocatalytic water splitting. Finally, the V-shaped crystal structure of monolayer \ce{Pt4I8} provides an excellent platform for alkali ion migration, with low diffusion barrier. In all, this study paves the way for the discovery of additional 2D materials with multifunctional properties, further advancing the development of the field of low-dimensional electronics.

%% classes. This should be given within the "acknowledgement"

\begin{acknowledgement}
	
This work is supported by the National Natural Science Foundation of China (22203026, 22403024, 22203025, 22103020 and 12174080), the Fundamental Research Funds for the Central Universities (JZ2024HGTB0162) and the Anhui Provincial Natural Science Foundation (2308085QB52). The computation is performed on the HPC platform of Hefei University of Technology.

\end{acknowledgement}


\begin{suppinfo}

Computational methods details of OER and HER; Structural information of 18 stable \ce{M4X8} and 24 unstable \ce{M4X8} compounds; Schematic illustration of 4 different magnetic orders of \ce{M4X8}; Phonon spectra and 300K AIMD simulations of 18 stable \ce{M4X8}; Phonon spectra and 300K AIMD simulations for 4 dynamically stable but dynamically unstable \ce{M4X8}; Variation of Young's modulus and Poisson's ratio with angle for 18 stable \ce{M4X8}; PBE band gaps and HSE band gaps of 18 stable \ce{M4X8} and their corresponding material types.

\end{suppinfo}


\bibliography{main}

\end{document}

\newpage
\section{Computational Methods Details}

To evaluate the electrocatalytic performance of the \ce{M4X8} monolayer, we calculated the oxygen evolution reaction (OER) and hydrogen evolution reaction (HER), which are the two half-reactions involved in the electrochemical water splitting. The OER (Equations (1)–(4)) and HER (Equations (5)–(6)) processes primarily follow the fundamental steps outlined below:

\begin{equation}
	*+\text{H}_{2}\text{O}\rightarrow\text{OH}^{*}+\text{H}^{+}+\text{e}^{-}
\end{equation}

\begin{equation}
	\text{OH}^{*}\rightarrow\text{O}^{*}+\text{H}^{+}+\text{e}^{-}
\end{equation}

\begin{equation}
	\text{O}^{*}+\text{H}_{2}\text{O}\rightarrow\text{OOH}^{*}+\text{H}^{+}+\text{e}^{-}
\end{equation}

\begin{equation}
	\text{OOH}^{*}\rightarrow*+\text{O}_{2}+\text{H}^{+}+\text{e}^{-}
\end{equation}

\begin{equation}
	*+\text{H}^{+}+\text{e}^{-}\rightarrow\text{H}^{*}
\end{equation}

\begin{equation}
	\text{H}^{*}+\text{H}^{+}+\text{e}^{-}\rightarrow*+\text{H}_{2}
\end{equation}
Here, * denotes the active sites on the electrocatalyst surface. The OER process involves a four-electron transfer, and its kinetic barrier is higher than the thermodynamic equilibrium potential, requiring a significant overpotential to drive the electrochemical reaction, which hinders the large-scale application of water electrolysis. The overpotential $\eta$ for the OER is estimated as follows:

\begin{equation}
	\eta=\frac{\max \left[\Delta G_{1}, \Delta G_{2}, \Delta G_{3}, \Delta G_{4},\right]}{e}-1.23+0.059 \times \mathrm{pH}
\end{equation}
Where $\Delta G_{i}$ represents the Gibbs free energy difference at each step. $e$ denotes electron volts.

For the catalytic performance of HER, the descriptor is typically based on the Gibbs free energy of hydrogen adsorption ($\Delta G$). The closer $\Delta G$ is to 0, the better the material's catalytic performance for HER. The change in Gibbs free energy, $\Delta G$, is calculated using the method proposed by N{\o}rskov et al., and is defined as follows:\cite{norskov2004origin}

\begin{equation}
	\Delta G=\Delta E+\Delta E_{\text{ZPE}}-T \Delta S+\Delta G_{\text{pH}}+\Delta G_{\text{U}}
\end{equation}
Here, $\Delta E$ corresponds to the energy difference between the free state and the adsorbed state of the reaction intermediate obtained from DFT calculations; $\Delta E_{\text{ZPE}}$ is the zero-point energy change, where the zero-point energy is taken from vibrational frequency calculations; $\Delta S$ is the entropy change, which can be obtained from the NIST database. Room temperature (T = 298.15 K) is used in the calculation. The free energy correction for pH is given by $\Delta G_{\text{PH}}$ = $k_B$T $\times$ pH $\times$ ln10. $\Delta G_{\text{U}}$ is calculated as $\Delta G_{\text{U}}$ = -eU, where U is the electrode potential relative to the standard hydrogen electrode (SHE).

\newpage
\begin{longtable}[htbp]{lccccccc}
	\caption{Lattice constants $a$ and $b$, magnetic ground state (as show in Figure \ref{fig:S_magset}), cohesive energy $E_{coh}$ (eV/atom), and phonon, molecular dynamics, and mechanical stability of \ce{M4X8}.}
	\label{table:S1} \\
	
	\hline
	System      & $a$    & $b$    & Magnetic ground State & $E_{coh}$ & Phonon & MD     & Mechanical\\  
	\hline
	\endfirsthead
	
	\hline
	System      & $a$    & $b$    & Magnetic ground State & $E_{coh}$ & Phonon & MD     & Mechanical\\  
	\hline
	\endhead
	
	\hline
	\endfoot
	
	\ce{Cr4Cl8} & 6.294  & 7.580  & AFM2     &  3.327  & $\surd$   & $\surd$   & $\surd$   \\
	\ce{Cr4Br8} & 6.760  & 8.101  & FM       &  2.915  & $\surd$   & $\surd$   & $\surd$   \\
	\ce{Cr4I8}  & 7.268  & 8.646  & FM       &  2.514  & $\surd$   & $\surd$   & $\surd$   \\
	\ce{Ni4F8}  & 5.533  & 5.697  & AFM1     &  3.529  & $\surd$   & $\surd$   & $\surd$   \\
	\ce{Ni4Cl8} & 5.997  & 7.337  & PM       &  3.273  & $\surd$   & $\surd$   & $\surd$   \\    
	\ce{Ni4Br8} & 6.349  & 7.675  & PM       &  2.948  & $\surd$   & $\surd$   & $\surd$   \\
	\ce{Ni4I8}  & 6.824  & 8.119  & PM       &  2.693  & $\surd$   & $\surd$   & $\surd$   \\   
	\ce{Cu4F8}  & 5.460  & 6.207  & FM       &  3.518  & $\surd$   & $\surd$   & $\surd$   \\  
	\ce{Cu4Cl8} & 6.256  & 7.720  & FM       &  2.699  & $\surd$   & $\surd$   & $\surd$   \\  
	\ce{Cu4Br8} & 6.634  & 8.120  & FM       &  2.388  & $\surd$   & $\surd$   & $\surd$   \\
	\ce{Mo4Br8} & 7.083  & 8.544  & FM       &  3.000  & $\surd$   & $\surd$   & $\surd$   \\
	\ce{Mo4I8}  & 7.477  & 8.945  & FM       &  2.647  & $\surd$   & $\surd$   & $\surd$   \\    
	\ce{Pd4Br8} & 6.707  & 8.700  & PM       &  2.532  & $\surd$   & $\surd$   & $\surd$   \\
	\ce{Pd4I8}  & 7.154  & 9.050  & PM       &  2.378  & $\surd$   & $\surd$   & $\surd$   \\   
	\ce{Ag4F8}  & 6.185  & 6.692  & AFM2     &  2.824  & $\surd$   & $\surd$   & $\surd$   \\    
	\ce{Ag4Br8} & 7.166  & 8.804  & PM       &  1.998  & $\surd$   & $\surd$   & $\surd$   \\
	\ce{Pt4Br8} & 6.703  & 8.959  & PM       &  3.018  & $\surd$   & $\surd$   & $\surd$   \\
	\ce{Pt4I8}  & 7.135  & 9.154  & PM       &  2.897  & $\surd$   & $\surd$   & $\surd$   \\
	
	\hline
	\multicolumn{8}{c}{Thermally Unstable} \\ 
	\hline
	
	\ce{Ru4Br8} & 6.796  & 7.920  & FM       &  3.132  & $\surd$   & $\times$  &     /     \\
	\ce{Rh4Cl8} & 6.351  & 7.196  & FM       &  3.315  & $\surd$   & $\times$  &     /     \\
	\ce{Tc4Cl8} & 5.580  & 7.587  & FM       &  3.495  & $\surd$   & $\times$  &     /     \\
	\ce{Tc4Br8} & 5.748  & 8.126  & AFM1     &  3.164  & $\surd$   & $\times$  &     /     \\
	
	\newpage
	
	%\hline
	\multicolumn{8}{c}{Dynamically Unstable} \\ 
	\hline
	
	\ce{Sc4F8}  & 6.312  & 7.424  & FM       &  5.435  & $\times$  &     /     &     /     \\
	\ce{Ti4F8}  & 5.225  & 5.532  & AFM1     &  5.323  & $\times$  &     /     &     /     \\
	\ce{V4F8}   & 5.996  & 5.740  & AFM1     &  4.551  & $\times$  &     /     &     /     \\
	\ce{Cr4F8}  & 5.816  & 6.253  & FM       &  3.998  & $\times$  &     /     &     /     \\
	\ce{Mn4F8}  & 6.210  & 6.127  & AFM1     &  4.150  & $\times$  &     /     &     /     \\   
	\ce{Cu4I8}  & 7.052  & 9.202  & PM       &  2.131  & $\times$  &     /     &     /     \\
	\ce{Zn4F8}  & 6.024  & 5.648  & PM       &  3.462  & $\times$  &     /     &     /     \\
	\ce{Mo4F8}  & 6.302  & 6.909  & AFM3     &  3.703  & $\times$  &     /     &     /     \\
	\ce{Mo4Cl8} & 6.817  & 8.190  & FM       &  3.382  & $\times$  &     /     &     /     \\
	\ce{Tc4F8}  & 4.969  & 6.487  & AFM3     &  4.364  & $\times$  &     /     &     /     \\
	\ce{Ru4F8}  & 4.522  & 6.690  & FM       &  4.176  & $\times$  &     /     &     /     \\    
	\ce{Ru4Cl8} & 6.449  & 7.652  & FM       &  3.459  & $\times$  &     /     &     /     \\
	\ce{Rh4F8}  & 5.224  & 7.734  & FM       &  3.880  & $\times$  &     /     &     /     \\
	\ce{Rh4Br8} & 6.812  & 8.284  & FM       &  3.025  & $\times$  &     /     &     /     \\
	\ce{Pd4F8}  & 5.604  & 7.732  & AFM3     &  3.290  & $\times$  &     /     &     /     \\    
	\ce{Pd4Cl8} & 6.394  & 8.531  & PM       &  2.786  & $\times$  &     /     &     /     \\
	\ce{Ag4Cl8} & 6.882  & 8.318  & AFM2     &  2.229  & $\times$  &     /     &     /     \\
	\ce{Ag4I8}  & 7.583  & 9.289  & PM       &  1.816  & $\times$  &     /     &     /     \\
	\ce{Pt4F8}  & 5.504  & 8.364  & PM       &  3.604  & $\times$  &     /     &     /     \\    
	\ce{Pt4Cl8} & 6.382  & 8.873  & PM       &  3.275  & $\times$  &     /     &     /     \\

	\hline
\end{longtable}

%\begin{comment}

\begin{figure}[H]
\centering
\includegraphics[width=0.8\textwidth]{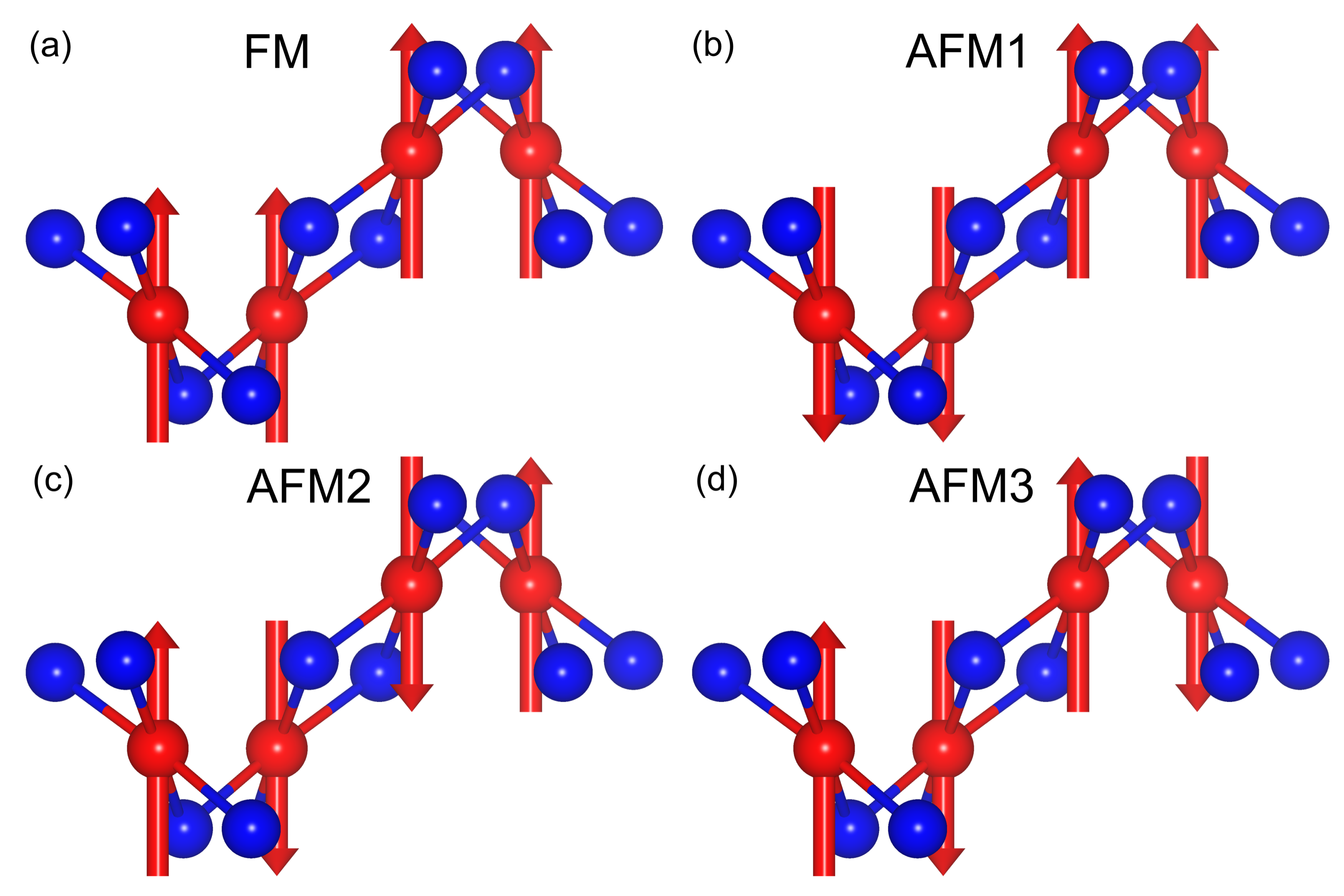}
\caption{Schematic illustration of four different magnetic orders of \ce{M4X8} monolayer. (a) FM. (b) AFM1. (c) AFM2. (d) AFM2.}
\label{fig:S_magset}
\end{figure}

\begin{figure}[H]
\centering
\includegraphics[width=0.8\textwidth]{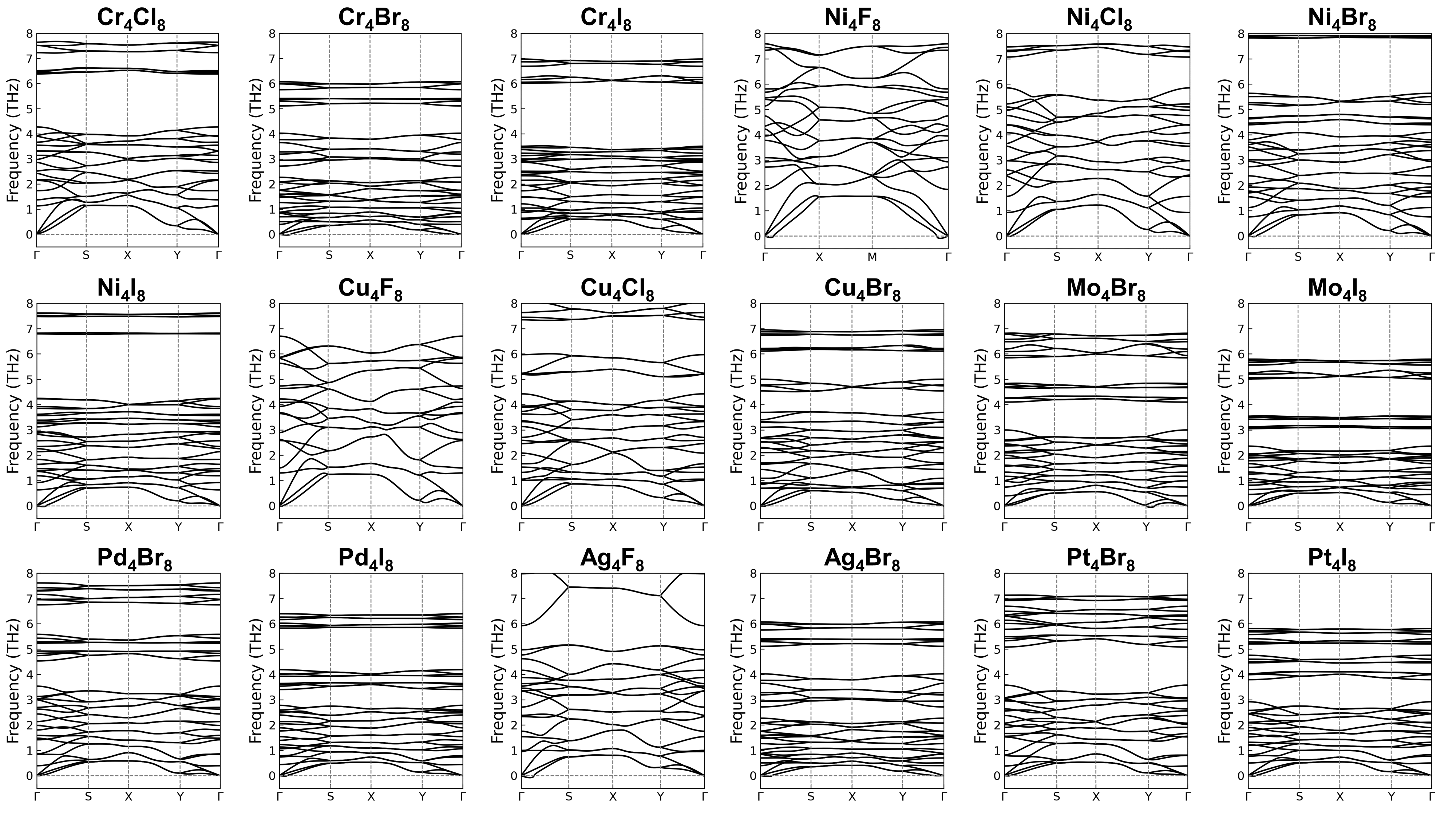}
\caption{Phonon spectra of the 18 \ce{M4X8} structures showing dynamical stability.}
\label{fig:S_donephonon}
\end{figure}

\begin{figure}[H]
\centering
\includegraphics[width=0.8\textwidth]{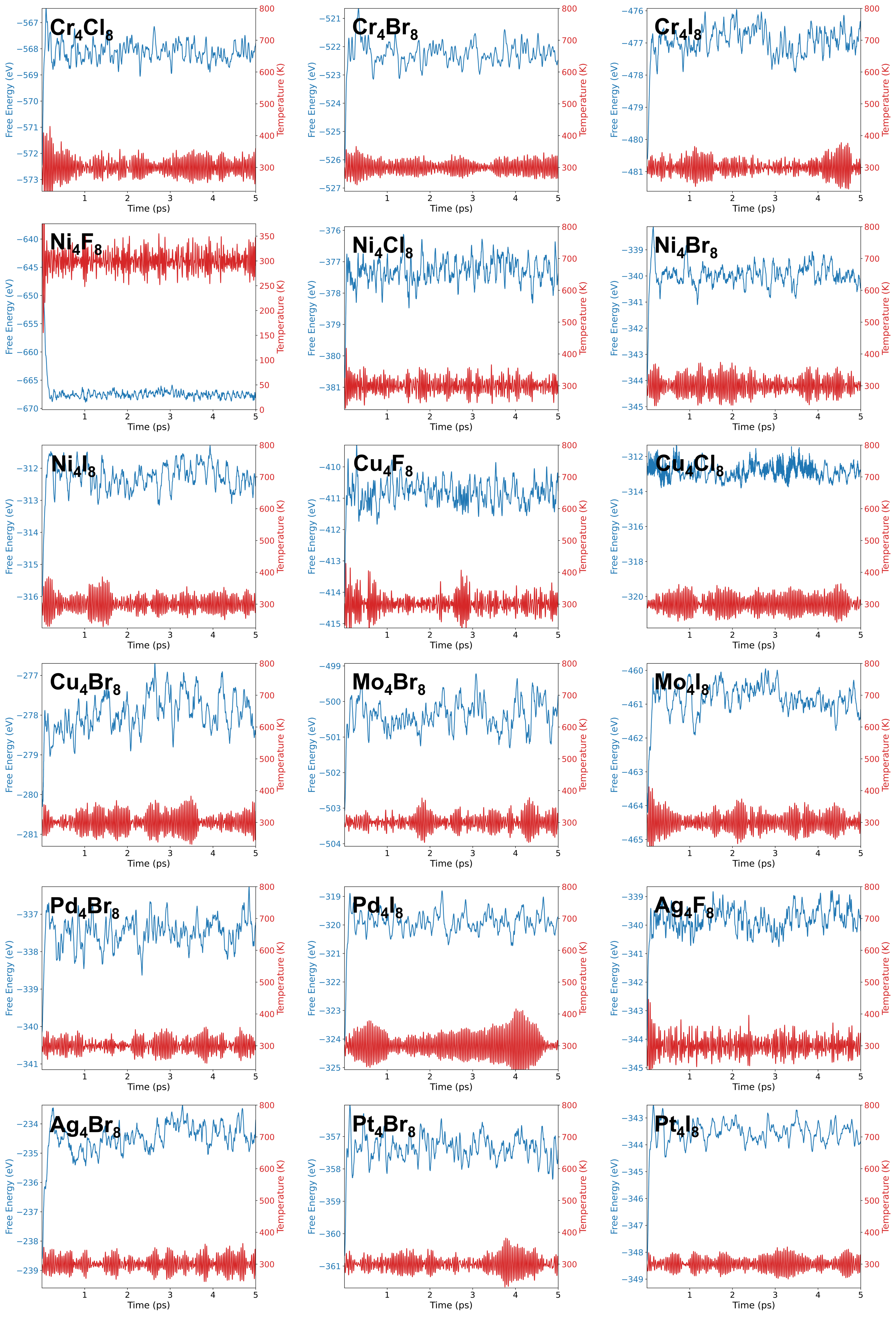}
\caption{Time-dependent relationship between free energy (blue) and temperature (red) for 18 thermally stable \ce{M4X8} structures at 300 K AIMD simulations.}
\label{fig:S_donemd}
\end{figure}

\begin{figure}[H]
\centering
\includegraphics[width=0.8\textwidth]{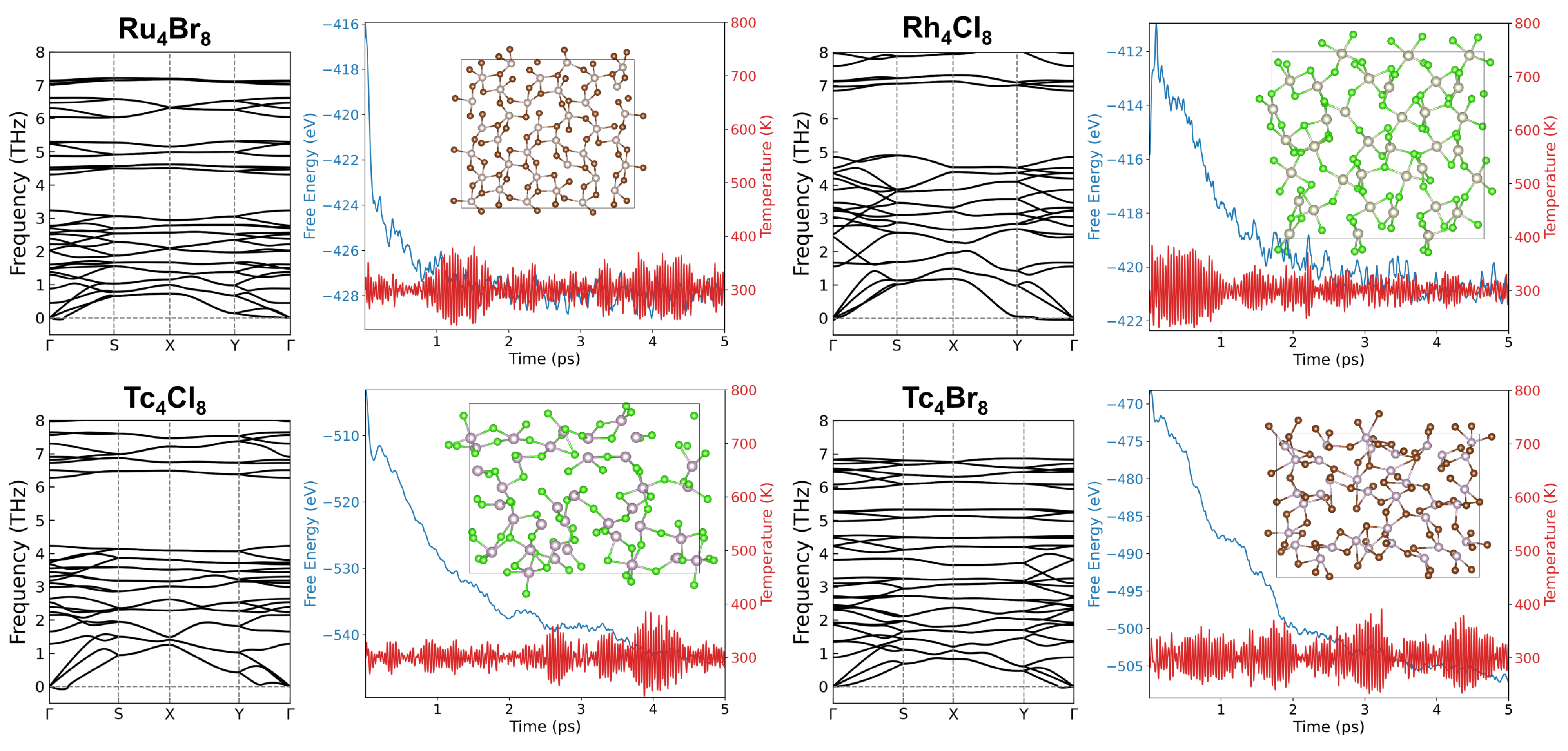}
\caption{Phonon spectra and 300 K molecular dynamics simulation results for four dynamically stable but dynamically unstable \ce{M4X8} structures. The illustration shows the supercell structure of \ce{M4X8} at 5 ps.}
\label{fig:S_phonon-done-md-fail}
\end{figure}

\begin{figure}[H]
\centering
\includegraphics[width=0.8\textwidth]{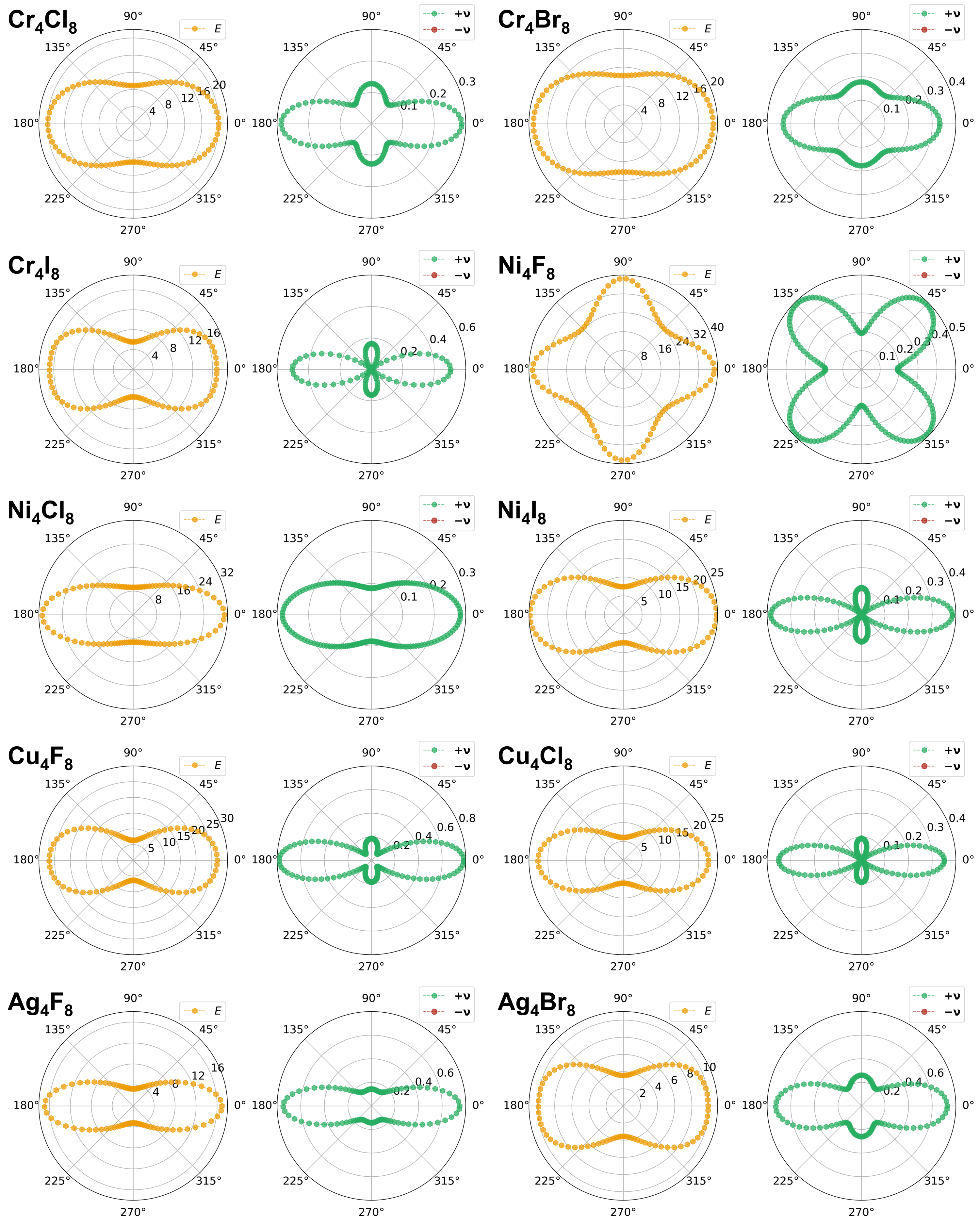}
\caption{Variation of Young's modulus and Poisson's ratio with angle for 10 \ce{M4X8} structures exhibiting positive Poisson's ratio.}
\label{fig:S_allppr}
\end{figure}

\begin{figure}[H]
\centering
\includegraphics[width=0.8\textwidth]{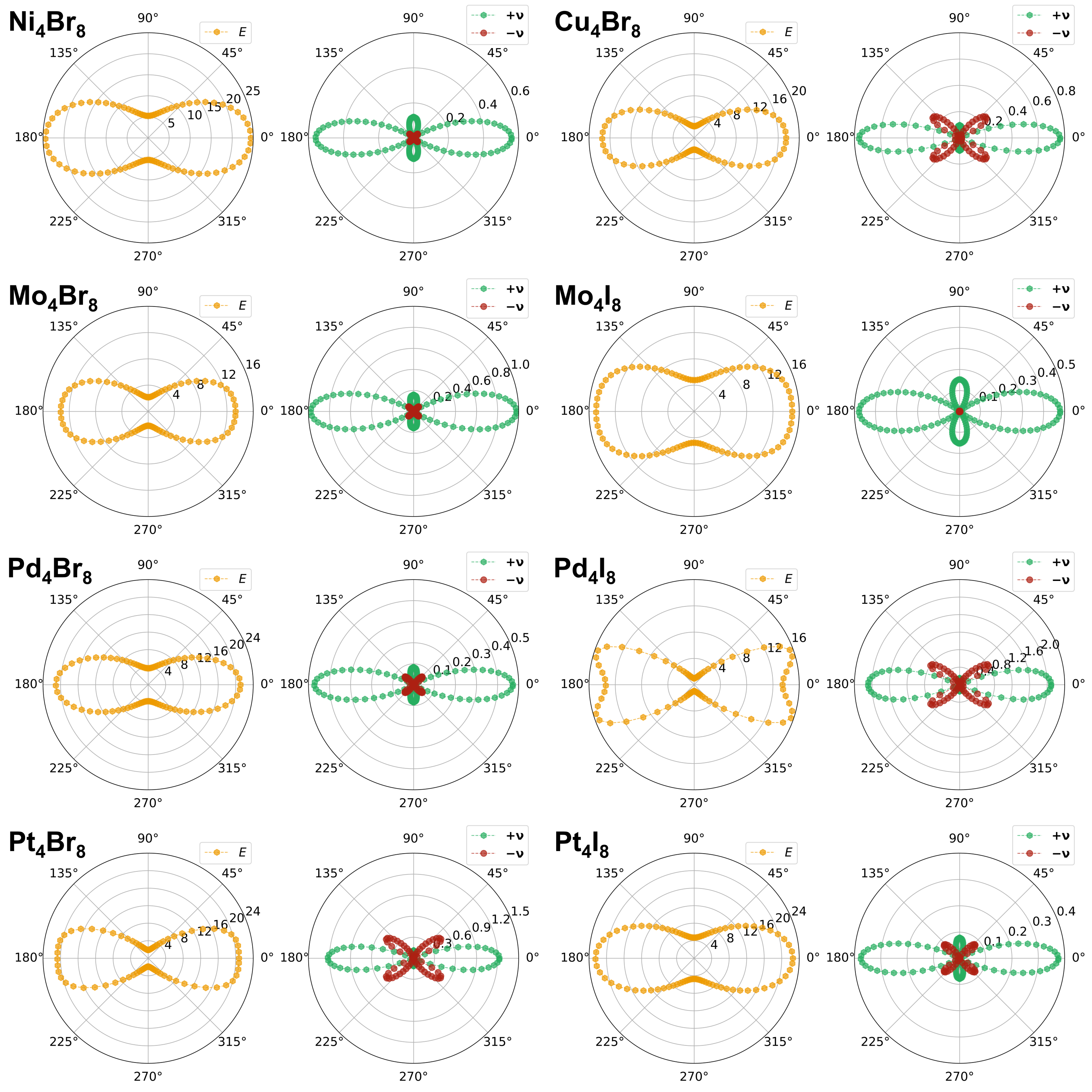}
\caption{Variation of Young's modulus and Poisson's ratio with angle for 8 \ce{M4X8} structures exhibiting negative Poisson's ratio.}
\label{fig:S_allnpr}
\end{figure}

\begin{table}[!htbp]
	\begin{center}
		\caption{PBE band gaps $Gap_{\text{PBE}}$ (eV) and HSE band gaps $Gap_{\text{HSE}}$ (eV) of 18 stable \ce{M4X8} structures and their corresponding material types.}
		\label{table:Electronic}
		\begin{tabular}{lccc}
			\hline
			System      & $Gap_{\text{PBE}}$ & $Gap_{\text{HSE}}$ & classification \\
			\hline
			\ce{Cr4Cl8} & 1.190 & 3.727 & AFMS   \\
			\ce{Cr4Br8} & 1.024 & 3.293 & HSC    \\
			\ce{Cr4I8}  & 0.833 & 2.568 & HSC    \\
			
			\ce{Ni4F8}  & 4.452 & 5.183 & AFMS   \\
			\ce{Ni4Cl8} & 0.663 & 3.127 & PMSCs  \\  
			\ce{Ni4Br8} & 0.624 & 2.652 & PMSCs  \\
			\ce{Ni4I8}  & 0.648 & 2.163 & PMSCs  \\
			
			\ce{Cu4F8}  & 0.608 & 4.573 & BMS    \\
			\ce{Cu4Cl8} & 0.179 & 2.437 & BMS    \\
			\ce{Cu4Br8} & 0.008 & 1.719 & BMS    \\
			
			\ce{Mo4Br8} & 1.324 & 3.192 & BMS    \\
			\ce{Mo4I8}  & 1.102 & 2.706 & BMS    \\
			
			\ce{Pd4Br8} & 1.175 & 2.520 & PMSCs  \\
			\ce{Pd4I8}  & 1.142 & 2.153 & PMSCs  \\
			
			\ce{Ag4F8}  & 0.433 & 2.770 & AFMS   \\
			\ce{Ag4Br8} & 0.000 & 0.000 & Metal \\
			
			\ce{Pt4Br8} & 1.477 & 2.797 & PMSCs  \\
			\ce{Pt4I8}  & 1.397 & 2.454 & PMSCs  \\
			\hline
		\end{tabular}
	\end{center}
\end{table}

%\end{comment}

\newpage

\bibliography{SI}